\documentclass[final]{agujournal2019}
\usepackage{apacite}
\usepackage{natbib} 
\usepackage{url} %this package should fix any errors with URLs in refs.
\usepackage{lineno}
\draftfalse

\journalname{JGR: Space Physics}

\begin{document}

\title{FORESAIL-1 cubesat mission to measure radiation belt losses and demonstrate de-orbiting}

\authors{M.~Palmroth\affil{1,2}, J.~Praks\affil{3}, R.~Vainio\affil{4}, P.~Janhunen\affil{2}, E.~K.~J.~ Kilpua\affil{1}, N.~Yu. Ganushkina\affil{2,5}, A.~Afanasiev\affil{4}, M.~Ala-Lahti\affil{1}, A.~Alho\affil{3}, T.~Asikainen\affil{6}, E.~Asvestari\affil{1}, M.~Battarbee\affil{1}, A.~Binios\affil{3}, A.~Bosser\affil{3}, T.~Brito\affil{1}, J.~Envall\affil{2}, U.~Ganse\affil{1},  H.~George\affil{1}, J.~Gieseler\affil{4}, S.~Good\affil{1}, M.~Grandin\affil{1}, S.~Haslam\affil{2}, H.-P.~Hedman\affil{7}, H.~Hietala\affil{4}, N.~Jovanovic\affil{3}, S.~Kakakhel\affil{7}, M.~Kalliokoski\affil{1}, V.~V.~Kettunen\affil{3}, T.~Koskela\affil{1,4}, E.~Lumme\affil{1}, M.~Meskanen\affil{2}, D.~Morosan\affil{1}, M.~Rizwan Mughal\affil{3}, P.~Niemel\"a\affil{3}, S.~Nyman\affil{3}, P.~Oleynik\affil{4}, A.~Osmane\affil{1}, E.~Palmerio\affil{1}, Y.~Pfau-Kempf\affil{1}, J.~Peltonen\affil{4}, J.~Plosila\affil{7}, J.~Polkko\affil{2}, S.~Poluianov\affil{6,8}, J.~Pomoell\affil{1}, D.~Price\affil{1}, A.~Punkkinen\affil{4}, R.~Punkkinen\affil{7}, B.~Riwanto\affil{3}, L.~Salomaa\affil{7}, A.~Slavinskis\affil{3,9}, T.~S\"antti\affil{7}, J.~Tammi\affil{7}, H.~Tenhunen\affil{7}, P.~Toivanen\affil{2}, J.~Tuominen\affil{7}, L.~Turc\affil{1}, E.~Valtonen\affil{4}, P.~Virtanen\affil{4}, T.~Westerlund\affil{7}}

\affiliation{1}{University of Helsinki, Department of Physics, Helsinki, Finland}
\affiliation{2}{Finnish Meteorological Institute, Space and Earth Observation Centre, Helsinki, Finland}
\affiliation{3}{Aalto University, School of Electrical Engineering, Espoo, Finland}
\affiliation{4}{University of Turku, Department of Physics and Astronomy, Turku, Finland}
\affiliation{5}{University of Michigan, Department of Climate and Space Sciences and Engineering, Ann Arbor, USA}
\affiliation{6}{University of Oulu, Space Climate Research Unit, Oulu, Finland}
\affiliation{7}{University of Turku, Department of Future Technologies, Turku, Finland}
\affiliation{8}{University of Oulu, Sodankyl\"a Geophysical Observatory (Oulu Unit), Oulu, Finland}
\affiliation{9}{University of Tartu, Tartu Observatory, T{\~o}ravere, Estonia}

\correspondingauthor{Minna Palmroth}{minna.palmroth@helsinki.fi}

\begin{keypoints}
\item FORESAIL-1 mission measures energetic electron precipitation and solar energetic neutral atom flux
\item We will demonstrate a cost-efficient de-orbiting and orbit manoeuvring technology without propellants
\item The goal of the mission is to contribute significantly to the sustainable utilisation of space
\end{keypoints}

\begin{abstract}

Today, the near-Earth space is facing a paradigm change as the number of new spacecraft is literally sky-rocketing. Increasing numbers of small satellites threaten the sustainable use of space, as without removal, space debris will eventually make certain critical orbits unusable. A central factor affecting small spacecraft health and leading to debris is the radiation environment, which is unpredictable due to an incomplete understanding of the near-Earth radiation environment itself and its variability driven by the solar wind and outer magnetosphere. This paper presents the FORESAIL-1 nanosatellite mission, having two scientific and one technological objectives. The first scientific objective is to measure the energy and flux of energetic particle loss to the atmosphere with a representative energy and pitch angle resolution over a wide range of magnetic local times. To pave the way to novel model - \textit{in situ} data comparisons, we also show preliminary results on precipitating electron fluxes obtained with the new global hybrid-Vlasov simulation Vlasiator. The second scientific objective of the FORESAIL-1 mission is to measure energetic neutral atoms (ENAs) of solar origin. The solar ENA flux has the potential to contribute importantly to the knowledge of solar eruption energy budget estimations. The technological objective is to demonstrate a satellite de-orbiting technology, and for the first time, make an orbit manoeuvre with a propellantless nanosatellite. FORESAIL-1 will demonstrate the potential for nanosatellites to make important scientific contributions as well as promote the sustainable utilisation of space by using a cost-efficient de-orbiting technology.

\end{abstract}

\section{Introduction}

Unprecedented numbers of new spacecraft are now being launched into Earth orbit to satisfy the growing demand from the scientific, commercial, and military sectors. Most of these new spacecraft need to survive in the radiation belts \citep[RBs;][]{vanAllen1959}, which are regions of trapped energetic charged particles. The RBs are critical in terms of space weather, as the radiation ages the spacecraft and deteriorates hardware. All new satellites contribute to the already existing large number of orbital debris, if they are not actively removed at the end of the mission.  This section introduces the state of the art in the three scientific and technological objectives of the FORESAIL-1 mission: measurements of energetic particle precipitation and solar energetic neutral atoms (ENAs), and de-orbiting technologies.

\subsection{State of the Art: Electron precipitation observations}

The RBs are produced by a complex balance of particle source and loss processes that vary both temporally and spatially \citep[e.g.,][]{Tverskoy1969, Walt1996, Chen2007, Shprits2008, Thorne2010}. Significant variations in electron fluxes occur over various time scales as a function of both energy and distance, driven by solar-magnetospheric interactions and internal magnetospheric processes \citep[e.g.,][]{Li1997, Elkington2003, Reeves2003, Shprits2006, Baker2008}. Effective losses from the outer radiation belts consist of 1) loss through the outer edge of the magnetosphere (magnetopause shadowing \citep[e.g.,][]{West1972, Ukhorskiy2006, Saito2010, Matsumura2011, turner2014}), 2) radial outward displacement of the electrons due to waves \citep{mann2016}, and weakening of the Earth's magnetic field (the $D_{st}$ effect (\citet{McIlwain1966, Kim1997, Millan2007}), and 3) wave-particle interactions resulting in scattering of particles into the loss cone \citep{Kennel1966, Thorne1971, Thorne1974}. There are no comprehensive estimates about which of these processes is most important during different conditions, while it is clear that particle losses play a central part in regulating the RBs. 

The wave-particle interactions leading to losses from the RBs vary on time scales ranging from 100 milliseconds to several minutes \citep{Millan2007}. Balloon experiments have historically been the earliest method to determine this loss category by measuring the X-rays from bremsstrahlung radiation induced by the interaction of precipitating particles with the neutrals in the upper atmosphere \citep{Barcus1973, Pytte1976}. Latest such observations are provided by the BARREL mission \citep{Millan2011, Woodger2015}. All balloon missions are constrained to balloon-reachable altitudes and thus only allow indirect observation of the precipitating particles. 

Energetic particle precipitation has also been observed from the ground, as precipitating electrons with energies over several tens of keV cause enhanced ionisation in the ionospheric D-region at an altitude of about 90 km. Relative ionospheric opacity meters (riometers) \citep{Hargreaves1969} are ground-based passive radars measuring the so-called cosmic noise absorption \citep{Shain1951}, which corresponds to absorbed radio wave power in the ionosphere resulting from enhanced D-region electron density. Recently, interferometric riometry has been developed to produce all-sky maps \citep[e.g.][]{McKayBukowski2015}. Ground-based observations of energetic electron precipitation can also be achieved using incoherent scatter radars, which can accurately measure D-region electron density down to about 70 km altitude \citep[e.g.][]{Miyoshi2015}. However, the intrinsically indirect ground-based observations do not allow inferring precipitating particle fluxes and energies, even using newly developed approaches such as spectral riometry \citep{Kero2014}. Hence having direct measurements of precipitating fluxes from space-borne instruments is critical for radiation belt loss studies.

One of the first satellite missions to study energetic electron precipitation was the Solar, Anomalous, and Magnetospheric Particle Explorer (SAMPEX, 1992 - 2012) used in a number of studies \citep{Lin2001, Tu2010, Nakamura2000}. DEMETER microsatellite observed electron fluxes at energies between 70 keV and 2.5 MeV with high energy resolution (256 channels) on a 700 km orbit \citep{Sauvaud2006}. These observations have been used to infer energetic electron precipitation \citep{Graf2009}, although DEMETER viewed primarily trapped particles. The main data set of direct measurements of precipitating energetic particles comes from the Medium Energy Proton and Electron Detector (MEPED) onboard NOAA/POES satellites \citep{Evans2000}. MEPED consists of two telescopes, the 0$^{\circ}$ telescope designed to measure precipitating particle fluxes and the 90$^{\circ}$ telescope for trapped particle fluxes, measuring electrons in three energy channels ($>$30 keV, $>$100 keV, and $>$300 keV) and protons in five energy channels ($>$30 keV, $>$80 keV, $>$250 keV, $>$800 keV, and $>$2.5 MeV). However, the NOAA/POES particle data suffer from two issues. First, the 0$^{\circ}$ telescope only partially views the bounce loss cone and does not offer any angular resolution inside its viewcone leading to poor pitch angle resolution. This leads to an underestimation of the precipitating fluxes \citep{Rodger2013}. Second, the electron channels are affected by proton contamination; partly corrected by a new data set by \citet{Asikainen2013}.

Particle precipitation is a key element in magnetosphere--ionosphere--thermosphere coupling, and therefore a crucial objective for research in numerical models, especially as there is an increasing demand for space weather forecasting capabilities. The first attempts to model precipitating particle fluxes relied on statistical patterns inferred from satellite observations. \citet{McDiarmid1975} produced a model for precipitating electron flux within 0.15--200~keV as a function of magnetic local time (MLT) and invariant latitude based on about 1100 passes of the ISIS~2 spacecraft. Using data measured by the Low Energy Electron experiment onboard the Atmosphere Explorer C and D satellites, \citet{Spiro1982} parametrised precipitating electron energy flux and average energy as a function of MLT, geomagnetic latitude, and geomagnetic activity measured by the $Kp$ and $AE$ indices. One of the reference models for auroral-energy electron precipitation is the \citet{Hardy1985} model, empirically derived by compiling several years of observations from the Defense Meteorological Satellite Program and Satellite (DMSP) Test Program spacecraft. The Hardy model is parametrised by the $Kp$ index, and it is still used to provide precipitation input in the 50~eV--20~keV energy range to state-of-the-art ionospheric models \citep[e.g.,][]{Marchaudon2015}. At higher energies (30~keV--1~MeV), the recent \citet{vandeKamp2016} model provides energy-flux spectra of precipitating electrons as a function of $L$ parameter and geomagnetic activity rendered with the $Ap$ index. This empirical model was obtained from a statistical analysis of 11 years of NOAA/POES energetic electron precipitation observations and is averaged across all MLTs in its present version.

\begin{figure}
  \centering
  \includegraphics[width=\textwidth]{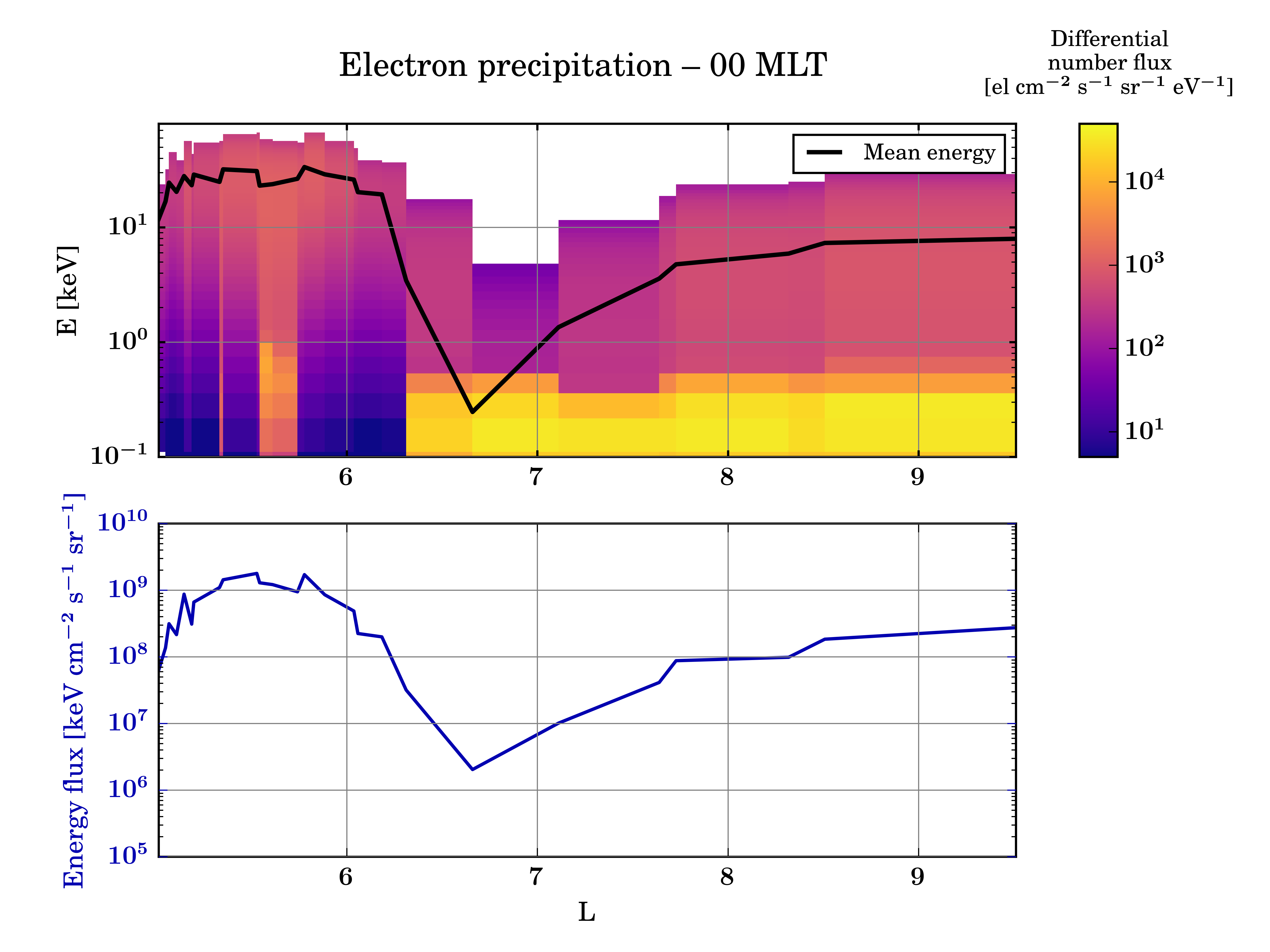}
  \caption{Preliminary results of Vlasiator modelling of electron precipitation. (top) Differential flux of precipitating electrons as a function of $L$ shell, in the same units as usually measured by telescopes onboard spacecraft. The black line indicates the mean precipitating energy. (bottom) Total precipitating energy flux as a function of $L$ shell.}
  \label{fig:Vlasiator_electrons}
\end{figure}

Modelling particle precipitation using first-principle models is not easy, given that many processes operating at spatial and temporal scales spanning many orders of magnitude are at play in the inner magnetosphere (energisation, and loss-cone scattering processes, among others). The emergence of global kinetic magnetospheric codes may enable addressing this issue in the near future. Recently, a preliminary run was performed using the Vlasiator code \citep{vonAlfthan2014,Palmroth2018}, in which electrons were added as a modelled species during a substorm-time, polar-plane global magnetospheric run throughout the magnetospheric simulation box. Figure~\ref{fig:Vlasiator_electrons} shows an example of these preliminary results of 0.1 $-$ 60~keV electron precipitation estimation obtained from the analysis of the nightside velocity distribution functions of electrons at a single time step of this simulation. The top panel shows the differential number flux of precipitating electrons as a function of $L$ shell (blue shading), as well as the mean precipitating energy (black line) in the same units as typical spacecraft data. The bottom panel shows the integral energy flux as a function of $L$. The integral energy flux was obtained by multiplying the differential number flux by the corresponding energies, and integrating across energies. The mean precipitating energy was calculated by dividing the integral energy flux by the total number flux (i.e., the differential number flux integrated across energies). The \citet{Hardy1985} model predicts a maximum integral energy flux of the order of 108 $-$ 109 keV cm$^{-2}$ s$^{-1}$ sr$^{-1}$ in the midnight sector, reached at geomagnetic latitudes comprised between 62$^{\circ}$ and 69$^{\circ}$ (translating into $L$ values between 4.5 and 7.8), depending on geomagnetic activity given by the $Kp$ index. The preliminary results from Vlasiator in Fig. \ref{fig:Vlasiator_electrons} are therefore in reasonable agreement with those values.
 
With the expanding human activity in space, it is increasingly important to measure precipitating particle fluxes \textit{in situ} and predict the precipitation by modelling, in order to understand the Earth's radiation environment. While previous satellite missions have provided a plethora of observations of the physical processes within the precipitation environment, none of the missions were designed specifically with a clear focus on precipitation. A number of new cubesat missions recently launched or being built focus on precipitation. These include, for example, the CSSWE mission \citep{Kohnert2011}, the ELFIN mission \citep{Shprits2018}, the Firebird mission \citet{Crew_Firebird_2016}, and the AMICal Sat mission \citep{Barthelemy2018}. FORESAIL-1 will be a complementary mission, improving the spatial and temporal resolution of precipitation that may be offered by these missions together.

\subsection{State of the Art: Solar energetic neutral atom observations }

The energy budgets of the solar corona and solar eruptions are major unsolved questions in solar physics. Coronal heating leads to an abundance of suprathermal particles in the corona. Suprathermal ions are important for estimating the energy budget of an eruption \citep[e.g.,][]{Emslie2004}, however, they do not produce measurable amounts of electromagnetic radiation, and thus their abundance is difficult to measure. Direct in-situ measurements of suprathermal ions will be provided by the recently launched NASA Parker Solar Probe mission, but only within the outermost reaches of the solar corona. During solar eruptions these suprathermal ions can be driven to participate in charge exchange processes with neutral atoms, resulting in the formation of solar energetic neutral atoms (ENAs). So far, ENAs have been measured only during one single event \citep{Mewaldt2009} with the IMPACT/LET instrument onboard the twin the Solar TErrestrial RElations Observatory (STEREO) spacecraft close to the beginning of the mission. Even these results may be questionable, as they have been disputed by \cite{Simnett2011}, who suggested the ENA observations could be explained by an earlier precursor event, detected as an electron burst.

\subsection{State of the Art: Space debris removal technologies}

The number of spacecraft in low-Earth orbit (LEO) is rapidly increasing, as technological and regulatory changes of launchers have allowed smaller satellites. These so-called nanosatellites typically do not have propulsion systems requiring bulky or volatile propellants for orbit control or de-orbiting, making them a significant source of future orbital debris.
Furthermore, if they cannot handle high-radiation environments, these nanosatellites will fail early, thus on one hand contributing to existing debris and on the other hand defeating potential plans for active deorbiting at the end of mission. The sustainable use of the near-Earth space has become of great interest \citep[e.g.,][]{Virgili2016}. To avoid low-Earth orbits becoming unusable in the future, also nanosatellites should be removed at end-of-life, otherwise the amount of space debris will increase exponentially due to collisions with bigger objects \citep{Kessler1978, Klinkrad1993, Bradley2009, Bonnal2013}, creating a potential danger to all later space missions. International standards have been developed to impose requirements on space missions to mitigate space debris production \citep[e.g.,][]{ESA_ADMIN_IPOL2014_2, ISO24113:2011}. Thus, it is crucial to develop robust instruments for both controlling the small satellite orbits as well as for removing them from orbit after the end of the mission.

Apart from technological solutions for reduction of space debris that are inherent to the satellite design, efforts for third-party orbit removal techniques are ongoing. Approaches include spacecraft that perform automatic rendez-vous, attachment and joint de-orbiting of larger space debris items. Using high-powered lasers (either ground- or satellite-based) to exert radiation pressure, or directly ablate the surface material of space debris (a so-called "Laser Broom") has been a research project in both civilian \citep{Bekey1997, Phipps2012} and military \citep{Campbell2000} projects. Compact propulsion methods possibly allowing de-orbiting of nanosatellites include miniaturised pulsed plasma and Hall-effect thrusters, which have reached commercial technological readiness, but still require propellants and a sizeable energy budget \citep{Tummala2017}. Photonic solar sails have been investigated for propellantless propulsion and used successfully for interplanetary missions \citep{Tsuda2013}, as well as multiple nanosatellite missions \citep{Lappas2011} with mixed success. Meanwhile, electric sails, in which electrically charged structures interact with the ion environment, have been proposed \citep{Janhunen2004}, and suitable packages have been implemented for nanosatellites, but successful experimental verification is still outstanding.

\section{Science goals}\label{txt:Objectives}

\subsection{Mission statement}
FORESAIL-1 is the first nanosatellite mission designed to measure the energy-dependent pitch angle spectra of the precipitating radiation belt particles, and solar ENA flux. Further, FORESAIL-1 will demonstrate the effectiveness of the plasma brake as a means of manipulating the spacecraft orbit in operation and lowering the spacecraft altitude to speed up de-orbiting at the end of the mission, thus addressing the sustainability of LEO space operations.

\subsection{Science objectives}

The FORESAIL-1 mission answers the following science questions: What are the pitch-angle and energy signatures of precipitation events as a function of MLT? How is the precipitation pitch-angle and energy distribution dependent on geomagnetic activity and driving from the solar wind? Thus, the FORESAIL-1 mission aims to perform precise directional measurements of electron and proton precipitation, as well as the energy spectrum and particle fluxes over a wide energy range (tens of keV to approximately one MeV). The spacecraft orbit shall drift in magnetic local time to allow determining precipitation as a function of MLT. The time resolution needs to be good enough to describe the lower bound on precipitation budget due to wave-particle interactions between chorus waves and electrons. Combining several measurements of at least three energy channels into a full pitch-angle resolution throughout the LEO region, with a time resolution of at least 15 s, will enable research of loss processes from the RBs. 

To understand the energy budget of solar eruptions, the second science goal of this mission is to measure solar ENAs. This requires novel observations in an energy range well exceeding the magnetospheric ENA range. For this purpose, we use the geomagnetic field as a filter of solar particles and measure the ENA flux, thus inferring the flux that originates from the solar direction. Solar ENAs can only be measured reliably at energies exceeding the ring-current ion energies.

\subsection{Technological objectives}

In addition to the science objectives outlined above, FORESAIL-1 has a technological objective to ensure the sustainable use of space and to set a precedent for maintaining clean orbits. The objective is to test the plasma brake technology and achieve at least a 100 km lowering of the spacecraft altitude at mid-to-high altitude LEO. The consequences of this lowering of the orbit are 1) the orbital drift of the mission allowing monitoring the science objectives as a function of MLT, and 2) demonstrating that the technology can be used to de-orbit spacecraft. We will observe efficiency and performance of the plasma brake during the experiment to determine general information about the orbit lowering process. The success of the plasma brake experiment (and thus the completion of the mission's sustainability goals) is dependent on the reliable operation of the avionics, making reliability a primary design driver for FORESAIL-1. Radiation effects are identified as a major potential source of failures, hence radiation hardening techniques are used in the design.

\section{Requirements}

The study of precipitating electrons from the RBs is intrinsically coupled to the characteristic energy ranges of the electron seed populations there. Scientific and operational requirements are as follows:
\begin{enumerate}
\item{NOAA/POES, for which the energy resolution is ($E_2$-$E_1$)/$E_1$ = 3 (where $E_1$ and $E_2$ are the upper and lower limits for consecutive integral channels), provides the lower reference bar in terms of energy resolution \citep{Evans}. The nominal energy resolution for FORESAIL-1 is 0.4 between the upper and lower limits of a channel.}
\item{The particle detector must have a sufficient discriminating ability between electrons and protons, such that the electron channel does not suffer from proton contamination. For lower energy channels there is no contamination, while in the higher energy channels we allow a small background, however, this should be so small that the electrons are discernible.}
\item{The orbit must drift to cover several MLTs.}
\item{For electron precipitation and solar ENA measurements, the orbital altitude should lie between 400 km and 800 km.}
\item{The electron pitch angle resolution should include at least three bins measured every 15 s.}
\item{The mission profile must allow for the use of the plasma brake for orbital and altitude control.}
\item{The mission profile must provide the ability for daily updates of measurements of orbital parameters to assess the effect of the plasma brake, once activated.}
\end{enumerate}

In order to achieve the above scientific requirements, the spacecraft spin axis must be oriented with 3$^{\circ}$ accuracy, with spin of 4.00 $\pm$ 0.04 rpm. Attitude determination system must supply the magnetic field vector with 1$^{\circ}$ accuracy and the satellite position must be known with 5-km accuracy. There must be a $\sim$1 kbs$^{-1}$ data downlink.

\section{Description of the Mission}
\subsection{General Concept}

FORESAIL-1 is a nanosatellite mission in LEO designed to answer the science objectives outlined in Section \ref{txt:Objectives}. The payload consists of a PArticle TElescope (PATE) and a Plasma Brake (PB). PATE will measure energetic electrons in the energy range $80-800\,$~keV as well as H\textsuperscript{+} ions (protons) and neutral atoms in the energy range $0.3-10\,$~MeV. PB consists of a tether that will be used to lower the spacecraft altitude. The spacecraft is constrained under the CubeSat 3U standard to fit the two payloads.

\subsection{Mission Timeline and Orbit}

The manufacturing of the FORESAIL-1 payload and spacecraft started in 2018 and the manufacturing and integration will continue throughout 2019 until launch. The spacecraft is scheduled to be launched in 2020 into a Sun-synchronous orbit at or lower than 700 km altitude. Available launch opportunities are sought in 2019. Following the successful launch of the mission, the 1-month commissioning phase is scheduled to start immediately. After the commissioning phase, the mission's primary science phase is scheduled for 4 months at the initial Sun-synchronous orbit. After the primary phase, the plasma brake will be applied to lower the spacecraft by more than 100 km such that 1) the plasma braking force is fully demonstrated and 2) from the lowered orbit the spacecraft will naturally de-orbit after 25 years. The lowering of the orbit will inject the spacecraft into a drifting polar orbit, allowing precipitation measurements in different MLTs. Following the successful orbital manoeuvring of the spacecraft, the science phase continues with detecting particle precipitation in the drifting orbit for at least 1$-$2 years. After this will be the ENA measurement phase. There is a possibility of an extended science phase that will be scheduled depending on the health of the spacecraft.

\subsection{Spacecraft Conjunctions}

FORESAIL-1 can be used in conjunction with various other spacecraft in order to determine the origin of precipitating particles observed. Spacecraft that can provide context to the FORESAIL-1 observations include the Solar and Heliospheric Observatory (SOHO), STEREO, the Advanced Composition Explorer (ACE), Wind, DSCOVR, the Geostationary Operational Environmental Satellites (GOES), and the Parker Solar Probe. These missions directly monitor solar wind conditions, coronal mass ejections and solar energetic particles.

The conditions encountered by FORESAIL-1 will depend strongly on the state of the other regions in the magnetosphere. Simultaneous observations from satellites such as Cluster, the Magnetospheric Multi-Scale mission, the Time History of Events and Macroscale Interactions during Substorms spacecraft (THEMIS), and the Geomagnetic Tail Lab (GEOTAIL) will be invaluable for understanding the global context in which the FORESAIL-1 measurements are made. In particular, when located in the relevant region, these spacecraft can monitor substorms occurring in the magnetotail and the associated fast earthward plasma flows which are also the sources of the energetic particle precipitations to be measured by FORESAIL-1, in addition to the RB source. They can also provide information about the wave activity in the magnetosphere, which will be key to interpreting the FORESAIL-1 observations. In the regions closer to Earth, data from the recently-launched Arase mission in the radiation belts will be of particular importance. Direct complementary observations to FORESAIL-1 will be provided by the POES satellites, which will provide precipitating particle data at similar energy ranges, however these data are often problematic and require corrections.

\subsection{Coordinated Ground-Based Observations}

Coordinated observation will also be possible with ground-based instrumentation. Whenever suitable conjunctions with riometer chains such as the Finnish chain operated by Sodankyl{\"a} Geophysical Observatory or the Canadian chain (NORSTAR) take place, it will be possible to compare energetic particle precipitation patterns to those inferred from cosmic noise absorption measurements. The special case of the Kilpisj{\"a}rvi Atmospheric Imaging Receiver Array (KAIRA) is of particular interest, as this instrument, which may be used as a multibeam, multifrequency riometer could allow to finely study energetic precipitation along the FORESAIL-1 overpass. Indeed, KAIRA can provide 1 s time resolution observations of cosmic noise absorption with beams of 10$^{\circ}$–30$^{\circ}$ width, depending on the considered frequency, which is accurate enough to study, e.g., individual patches of pulsating aurora with KAIRA \citep{Grandin2017}, suggesting that it may be possible to study mesoscale ($<$100 km) structures in energetic precipitation using FORESAIL-1 and KAIRA conjunctions.

During conjunctions, it can also prove valuable to combine FORESAIL-1 data with observations of other space weather manifestations. For instance, auroral precipitation can be inferred from observations by all-sky imagers such as the MIRACLE network in Fennoscandia, and the geomagnetic context of FORESAIL-1 measurements can be given more accurately during conjunctions with ground-based magnetometer networks as well as by making use of polar cap convection maps derived from Super Dual Auroral Radar Network observations (SUPERDARN). Finally, FORESAIL-1 precipitation data could prove useful in studies focusing on electron density enhancements in the ionosphere using, e.g., ionosondes or incoherent scatter radars such as the European Incoherent Scatter radars (EISCAT). The latter ones also enable the study of effects such as ionospheric Joule heating or ion outflow, as they measure ion and electron temperatures and ion line-of-sight velocity, and when in a specific configuration they also allow estimating electric fields \citep{Nygren2011}.

\section{Payloads}

The FORESAIL-1 mission will carry two science payloads, the Particle Telescope (PATE) and the Plasma Brake (PB).

\subsection{Particle Telescope (PATE)}

PATE measures energetic electrons, H\textsuperscript{+} ions (protons) and neutral H atoms. The targeted energy range of hydrogen measurement is $0.3-10\,$MeV, which covers the energies above the typical ring-current proton energies. This is to avoid the neutral hydrogen background originating from the interaction of the ring current with the geocorona. The primary energy range for electrons is $80-800\,$keV. In addition, the instrument is sensitive to electrons at higher and lower energies in channels, where reliable particle identification cannot be performed, but especially the high-energy integral channel is still valuable since the contamination from heavier species comes only from relativistic protons, which have low fluxes compared to relativistic electrons in typical conditions.

\begin{figure}
\centering
\includegraphics[width=0.8\textwidth]{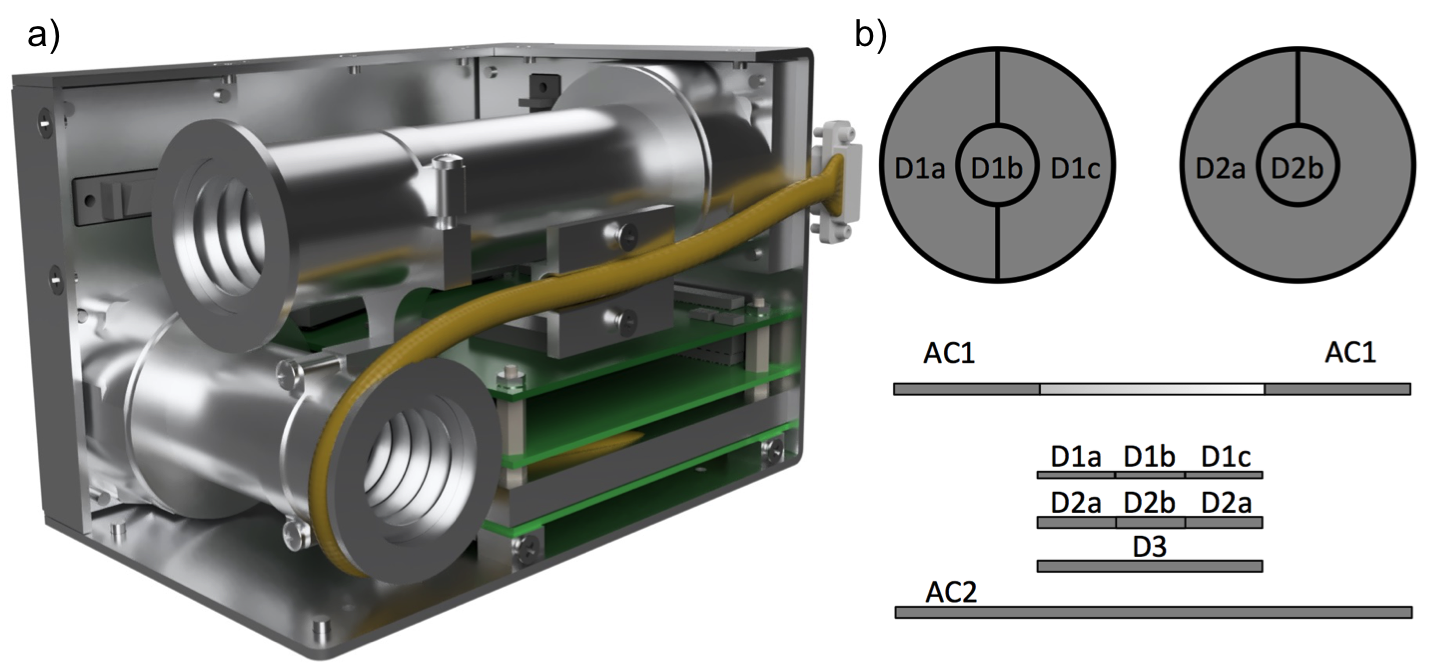}
\caption{a) Mechanical design concept of PATE. b) Schematic of the anti-coincidence (AC) and main detector (D) stack of each telescope.}
\label{fig:PATE_mechanical_design}
\end{figure}

PATE consists of two particle telescopes with identical stacks of detectors (see Fig.\ \ref{fig:PATE_mechanical_design}). The longer Telescope 1 is directed along the long axis of the spacecraft, that is, perpendicular to the rotation axis, and thus, it scans the directions in a plane perpendicular to the rotation axis of the spacecraft. The shorter Telescope 2 is directed along the rotation axis, so it can maintain a stable orientation. When the rotation axis is pointed towards the Sun, the telescope is able to measure the neutral hydrogen flux from the Sun. Note that the instrument itself does not determine the hydrogen charge state but relies on the geomagnetic field as a rigidity filter and on the measurement of angular distribution to disentangle the neutral flux from the solar direction. The motivation for the use of longer collimator in Telescope 1 is to improve the pitch angle angular resolution to better than 10 degrees for the scanning telescope.

Both telescopes have a mechanical collimator defining the aperture, consisting of an aluminum tube housing 18 (12) ($500\,\mu$m Al + $500\,\mu$m Ta) collimator rings in Telescope 1 (Telescope 2), followed by a stack of silicon detectors, D1, D2, and D3, measuring the energy losses of the particles in adjacent layers. The measured signals allow the determination of particle energy and the identification of particle species (electron / H). The thicknesses of the D detectors are $20\,\mu$m, $350\,\mu$m, and $350\,\mu$m, respectively. The D1 detectors have a bias voltage of 5 V, while the other ones are biased at 70 V. D1 and D2 detectors are segmented so that the central elements have diameters of 5.2 mm while the total active-area diameters of both are 16.4 mm. D3 has a single active area of 16.4 mm diameter. Both detector stacks are covered at the bottom of the collimator by two thin (nominally $0.5\,\mu$m each, 1 mm apart) Ni foils preventing low-energy ($<250\,$keV) ions and soft ($<500\,$eV) photons from entering the detector stack. Each aperture is limited from above by an annular anti-coincidence detector AC1 ($300\,\mu$m thickness) with a circular hole of 14.0 mm diameter in the middle and and an outer active-area diameter of 33.8 mm. Another single-element circular anti-coincidence detector AC2 ($350\,\mu$m thickness, 33.8 mm active-area diameter) at the bottom of the stack flags particles penetrating the whole D detector stack. Note that while the AC2 is operated in anti-coincidence with the D detectors for the nominal energy range of the instrument (particles stopping in the D stack), PATE still analyses the D detector pulse heights for particles triggering AC2 but not AC1 to provide integral flux channels above the nominal energy ranges. The distances from upper surface to upper surface in the detector stack AC1--D1--D2--D3--AC2 are 2.5 mm, 2.0 mm, 2.5 mm, and 2.5 mm. The lower Ni foil is 3.2 mm above the upper surface of AC1.

Signal processing is based on continuous sampling and digitization of the analog signals and on digital filtering and pulse height analysis. The signal processing board contains 16 Analog to Digital Converters (ADCs) and a \emph{Microsemi ProAsic3L} Field Programmable Gate Array (FPGA) handling the signal processing for both telescopes. The signal sampling rate in each ADC channel is 10 MHz, and the digital streams are processed by the FPGA, which is running at 40 MHz. The logic analyzes the incoming digital data streams, detects pulses, and identifies the particle for each valid coincidence event, counting and rejecting from further analysis any events not matching validity criteria. Valid particle events are then counted in separate counters based on their detection time, species and measured energy, forming the bulk of the science data of the instrument. The electron (hydrogen) spectrum consists of seven (ten) energy channels, log-spaced in measured energy.

\subsubsection{Instrument Performance}
\begin{figure}
\centering
\includegraphics[width=0.7\textwidth]{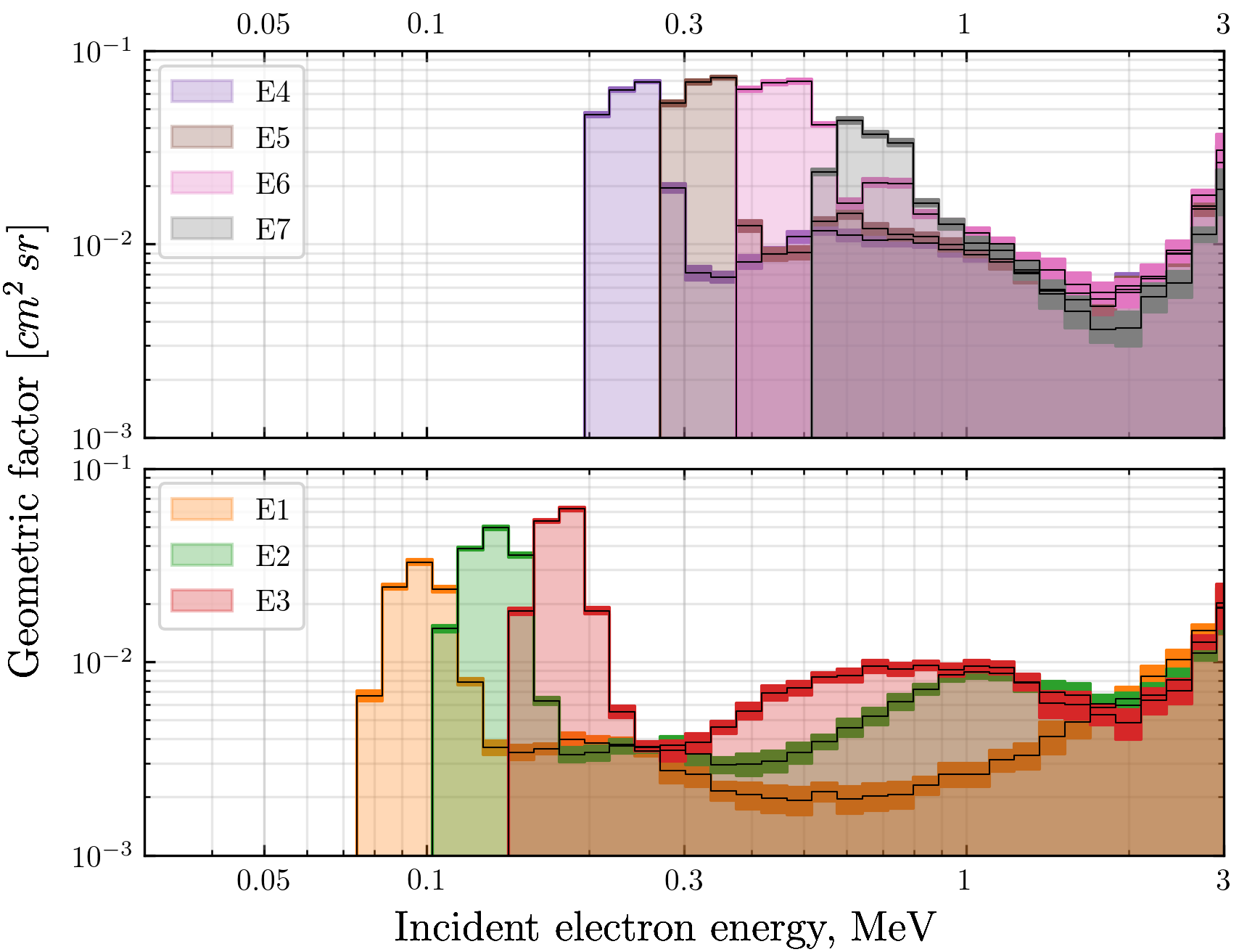}
\caption{The geometric factor of PATE as a function of energy for the seven electron energy channels within the nominal energy range (80--800 keV), simulated using GEANT4 \citep{GEANT4}. The range above 800 keV is additionally covered by a penetrating particle flux channel.}
\label{fig:PATE_efficiency}
\end{figure}

The performance of the PATE electron measurement has been simulated using GEANT4 \citep{GEANT4}. The simulation is performed for an isotropic electron distribution launched from a (15-cm radius) spherical surface surrounding PATE. The simulated pulse heights of all D-detector signals are analysed to separate electrons and hydrogen (ions/ENAs) and to measure particle energies, as in the FPGA-based on-board analysis. Particles producing a hit (i.e., an energy deposit $>50\,$keV) only in D2 are identified as electrons and particles producing a hit only in D1 are identified as hydrogen. Hit levels can be set freely in the logic, but values lower than 50 keV in D1a, D1c, AC1 and AC2 (see Fig.\ \ref{fig:PATE_mechanical_design}) should not be used as the simulated RMS noise levels in those pads are around 9--14 keV. While electrons are able to produce hits in D1 as well, there is only a minor level of electron contamination in the nominal hydrogen energy channels, which require the energy deposit in D1 to exceed 110 keV. If more than two adjacent D layers detect a pulse, the Delta E -- E method is used for clean species separation. The geometric factor of the seven electron channels as a function of electron energy for (the shorter) Telescope 2 is shown in Fig. \ref{fig:PATE_efficiency} \citep{Oleynik2019}. The high-energy tails of the response functions are due to the inevitable scattering of electrons off the detectors and other structures inside the telescope, which prevents the full energy of the electron to be absorbed in active detector elements. The internal energy resolution of the instrument is much better for hydrogen than for electrons and the response functions are close to boxcar type within the nominal energy range.

The geometric factors of both telescopes are mostly determined by the upppermost collimator ring (with an inner diameter of 21.5 mm) and the hole of the AC1 detector (diameter of 14 mm), which are at a vertical separation of 12.0 cm (7.0 cm) in Telescope 1 (Telescope 2). The nominal value of the geometric factor is 0.037 cm$^2\,$sr (0.11 cm$^2\,$sr) for Telescope 1 (Telescope 2), but especially electrons have somewhat lower values (see Fig.\ \ref{fig:PATE_efficiency}) due to scattering off the Ni foils and the D1 detector, causing trajectories to miss the D2 detector. The nominal angular widths of the acceptance cones (half width at half the on-axis value) for the two telescopes are 4.6$^\circ$ and 7.9$^\circ$, respectively. The instrument can also be operated in a mode where only the central segments of the D1 and D2 are included in the coincidence logic while the rim areas are logically included in the anti-coincidence. This allows to decrease the geometric factors of the telescopes by a factor of about 7. Any detector element can also be switched off from the logic entirely. 

\subsubsection{Mass, power and telemetry budgets}
The mass of PATE is $1.2\,$kg, consisting of the instrument box and mechanical support structures (435 g), detector and pre-amplified board housings (290 g), the collimators (180 g), cables (115 g), and the rest (180 g), including the back-end electronics stack. The power budget for PATE is $2.5\,$W, half of which is consumed by the FPGA, with an additional margin of 20\%, mainly required to account for the final FPGA power consumption.

The telemetry budget of PATE is summarized in Table \ref{tab:PATE_Telemetry_budget}. Spectral resolution of the data products for both electrons and hydrogen is on average $\Delta E/E \approx 40\%$ within the nominal energy ranges, which means that the spectral counter data consists of eight (ten) differential and one (two) integral channels for electrons (hydrogen). The rotation period,  nominally 15 s, of the satellite equals the time resolution of the instrument. This main time frame is further broken in 36 angular sectors for the rotating telescope to provide the angular distribution measurement. Both telescopes deliver, in addition to the spectral counter data, also pulse height data samples, which allow an accurate in-flight calibration and health monitoring of the detection system.

\begin{table}
\centering
\begin{tabular}{lcc}
Data source                & Data rate [bit/s]  & Data amount per day [kiB]   \\
\hline
Rotating telescope         & 993                & 10500                       \\
Solar pointing telescope   & 60                 & 633                         \\
Housekeeping               & 13                 & 135                         \\
\hline
Total                      & 1066               & 11268 \\
\end{tabular}
\caption{Summary of the PATE telemetry budget during science operations.}
\label{tab:PATE_Telemetry_budget}
\end{table}

\subsection{Plasma Brake (PB)}

\subsubsection{Operating Principle}

\begin{figure}
\centering
\includegraphics[width=\textwidth]{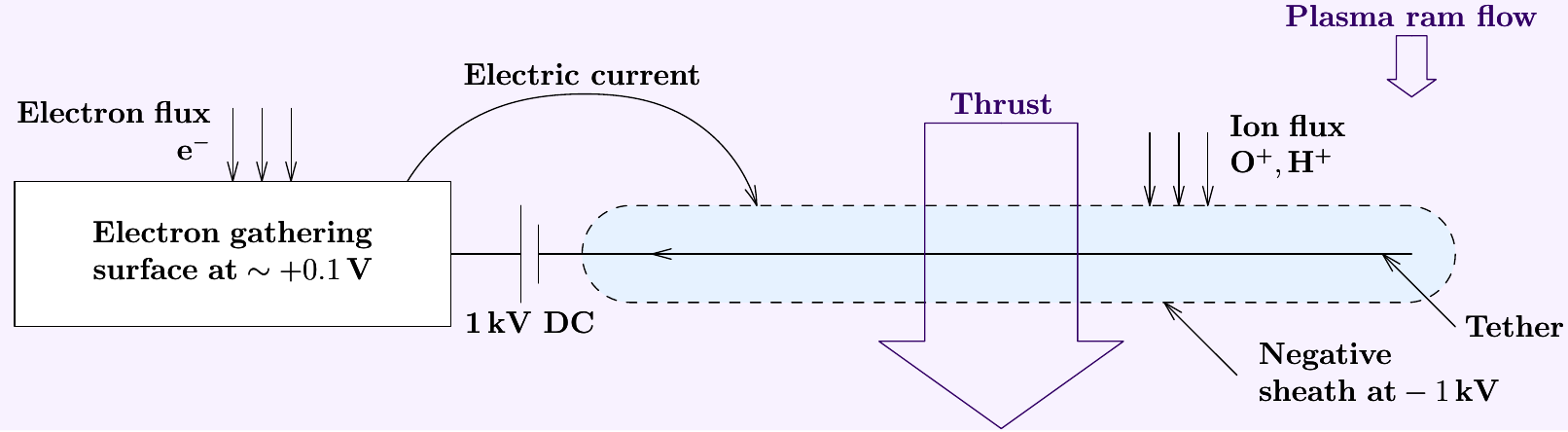}
\caption{Operating principle of the PB: an electric current system and a net thrust exerted on the negative tether by the plasma ram flow.}
\label{fig:PB_operating_principle}
\end{figure}

The Plasma Brake instrument is designed to measure the Coulomb drag, i.e., the braking force caused by the ionospheric plasma ram flow to an electrically charged tether (Figure \ref{fig:PB_operating_principle}). When interacting with the surrounding plasma, the negative tether gathers positive ions, which tends to neutralize the tether. To maintain the charge, the tether is connected by a high-voltage power system to a conducting surface (deployable booms for FORESAIL-1) that acts as an electron sink to close the current system through the plasma \citep{Janhunen2010}. The braking force can be measured in two modes. One is to monitor the system spin rate while charging the tether synchronously with the tether rotation (PB Measurement). The other is to maintain a constant charging and monitor the spacecraft velocity and orbital elements (PB Mode).

We employ $-1$ kV voltage, which is the maximum without risking electron field emission from micrometeoroid-struck parts of the tether wires. At this voltage, the expected nominal plasma brake thrust per tether length is 58 nN/m when the tether is perpendicular to the ram flow. This value is obtained by using Equation 1 of \citet{Janhunen2014} and assuming plasma density of $3\cdot 10^{10}$ m$^{-3}$, mean ion mass of 10 proton masses, ram flow speed of 7.8 km/s and tether's effective electric radius of 1 mm. The tether's collected ion current is small, nominally 30 $\mu$A for a 300 m long tether.

\subsubsection{Design}

\begin{figure}
\centering
\includegraphics[width=0.35\textwidth]{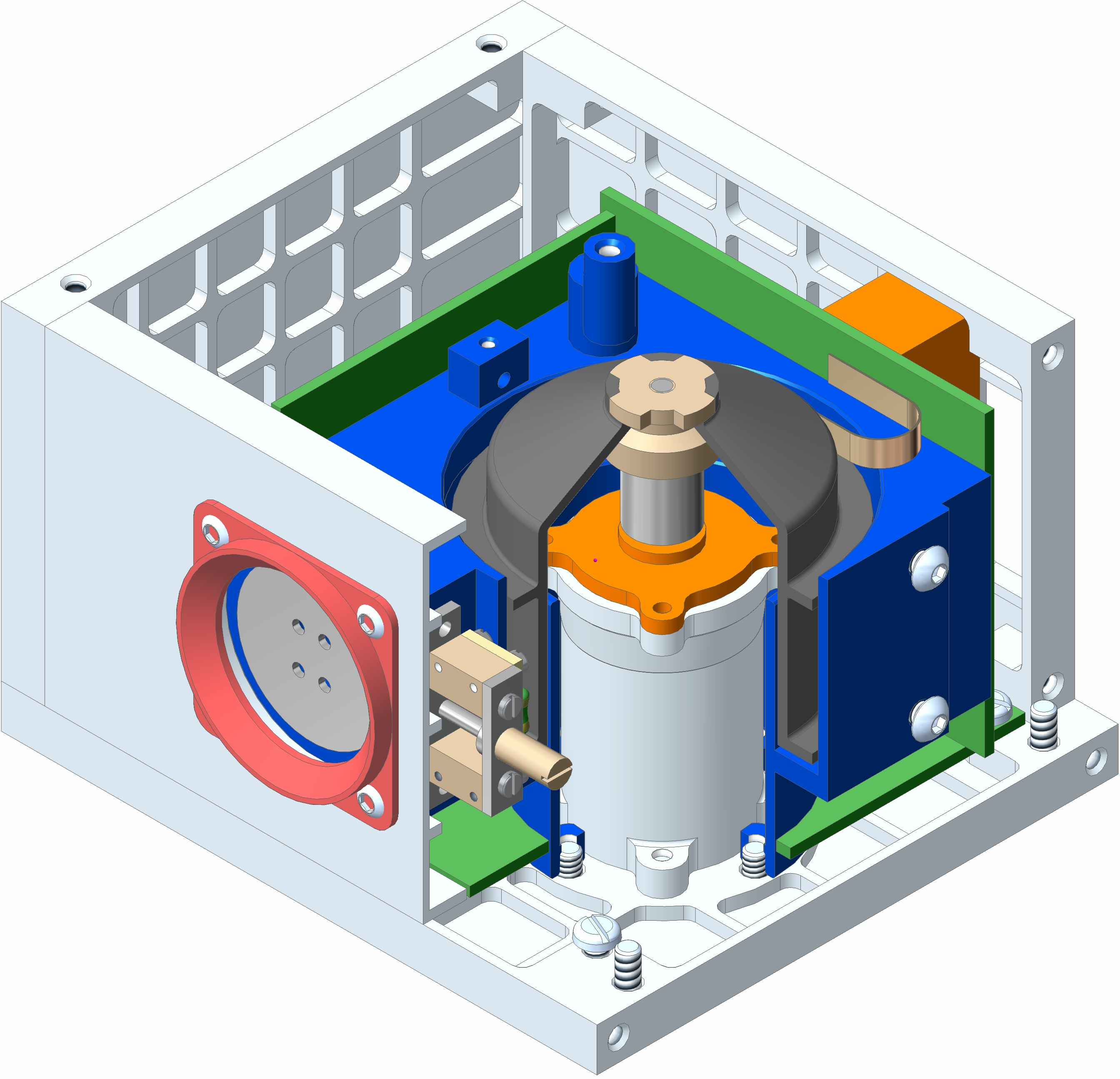}
\caption{PB payload structure.}
\label{fig:PB_structure}
\end{figure}

The tether is deployed from a chamber (blue) by the centrifugal force affecting the tether tip mass (gray button inside the red collar) (Figure \ref{fig:PB_structure}). The mechanical interface through the satellite side panel is provided by an anti-static collar (red) to avoid triple junctions of plasma, insulator, and high voltage tether. The tip mass is locked during the launch by two launch locks located on opposite sides of the tether chamber opening. The tether reel (dark gray) and the stepper motor that turns it (orange) are nested inside the tether chamber. The tether high-voltage contact is through the conducting reel and the slider (copper brown) attached straight to the high-voltage converter (orange) board behind the tether chamber.

The PB tether is made of thin metallic wires that are periodically bonded to each to withstand the micrometeoroid and space debris collisions \citep[e.g.,][]{Seppanen2011}. The single wire thickness is a few tens of micrometers to minimise the ion current to the tether and thus the mass of the power system and size of the electron gathering surface.

The nominal FORESAIL-1 tether length is 300 meters requiring $\sim$20~Nms of the total angular momentum to deploy it. The initial angular speed of 180 $^{\circ}$/s is sufficient to provide a centrifugal force of 0.4~cN which would safely pull out a tip mass weighing 2.5 g without breaking the tether. The angular momentum is provided by several consecutive satellite spin-up and reel-out maneuvers to compensate for the decreasing spin rate associated with the increasing moment of inertia. Since magnetorquers are used for attitude control, the angular acceleration is low to avoid the tether winding around the satellite. It also indicates that a considerable amount of time is needed to provide the angular momentum. However, after measuring the Coulomb drag force with a few tens of meters of the tether, the force can be used to spin up the satellite by modulating the tether voltage in synchronization with the rotation (charging downstream to increase the spin rate). After deploying the tether and taking PB measurements, the PB mode will start by continuously charging the tether which in turn will lower the orbit and start the satellite drift in MLT. When reaching $\sim$600-km altitude and gaining a sufficient drift in MLT, the satellite will be prepared for PATE observations -- the tether will be discarded to allow the spin axis being pointed towards the Sun. It can be done by attempting to reel in the tether which might break because it would be partially broken by micrometeoroids. A broken tether would deorbit in a few months thanks to its large area/mass ratio. If the tether does not break, the attitude control system and/or PB itself will have to provide an angular momentum to compensate for an increasing spin rate.

\subsubsection{Mass, Power and Telemetry Budgets}

The mass of the PB is 0.6 kg including a margin of 25\%. The structure (frame, tether chamber, and motor mounting shaft) contributes 67\% to the mass budget. The size of the system is 67$\times$84$\times$96~mm. For the two measurement modes of the PB, the power budget for PB is 600 mW. For the tether reeling, the motor and the controller consumes 7 W continuously. In case the reel-out power cannot be provided continuously by the spacecraft, the operation can be done in stages. The telemetry budget of the PB is summarised in Table
\ref{tab:PB_Telemetry_budget}. To reduce the overall telemetry budget, the long-duration routine PB mode uses a lower telemetry rate. Frequent housekeeping data are not required because in the PB mode, altitude change over weeks to months will show the success of experiment.

\begin{table}
\centering
\begin{tabular}{lcc}
Mode          & Data rate [bit/s]  & Data amount per day [kiB]   \\
\hline
Reel-Out/In         & 128                & 1350                       \\
PB Mode   & 19                 & 200                       \\
PB Measurement               & 256                 & 2700      \\
Standby & 64 & 675
                   \\
\hline\\
\end{tabular}
\caption{Summary of the PB telemetry budget for the operation modes.}
\label{tab:PB_Telemetry_budget}
\end{table}

\section{Spacecraft platform}

\begin{figure}
\centering
\includegraphics[width=\textwidth]{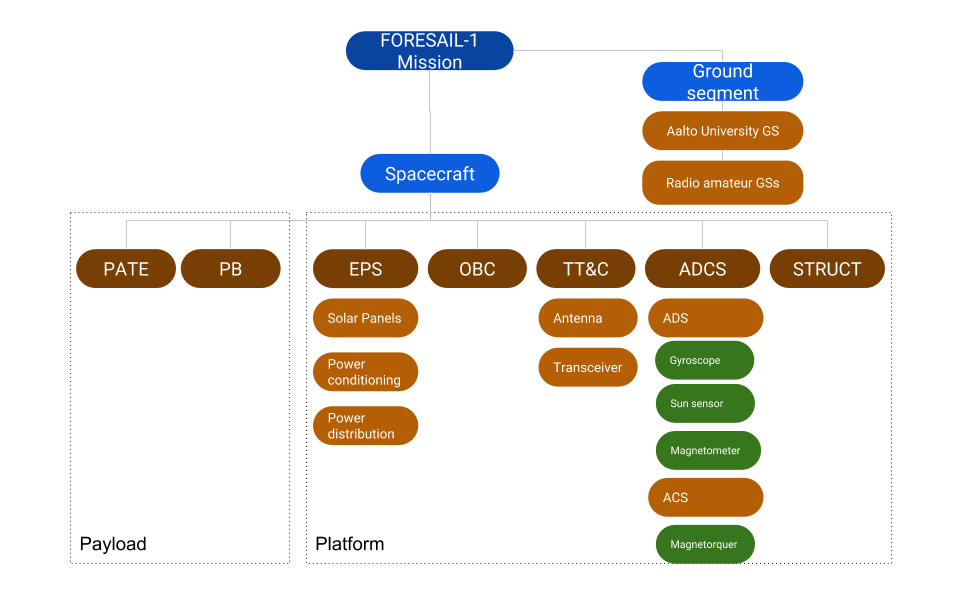}
\caption{FORESAIL-1 mission product tree.}
\label{fig:FS1_product_tree}
\end{figure}

The platform has been designed to accommodate the payloads and to provide data, power and mechanical interfaces. The overall mission tree for the FORESAIL-1 is presented in Figure \ref{fig:FS1_product_tree}. Since one of the key technological drivers for the mission is reliability, the avionics stack has several designs to this end. The avionics stack is enclosed in a vault providing substantially better shielding than what is typically seen on CubeSats (around 4 mm equivalent aluminium, instead of the more typical 2 mm), thus enhancing system tolerance to total dose. Single-event effects will be mitigated using dual cold redundancy, hardware overcurrent protection, minimization of the software footprint, and systematic data integrity checks. Finally, the FORESAIL-1 components will be submitted to radiation test campaigns; the radiation response data will be made available publicly in order to benefit from the broader field of (small) satellite technology and help other designers addressing this issue.

\begin{figure}
\centering
\includegraphics[width=\textwidth]{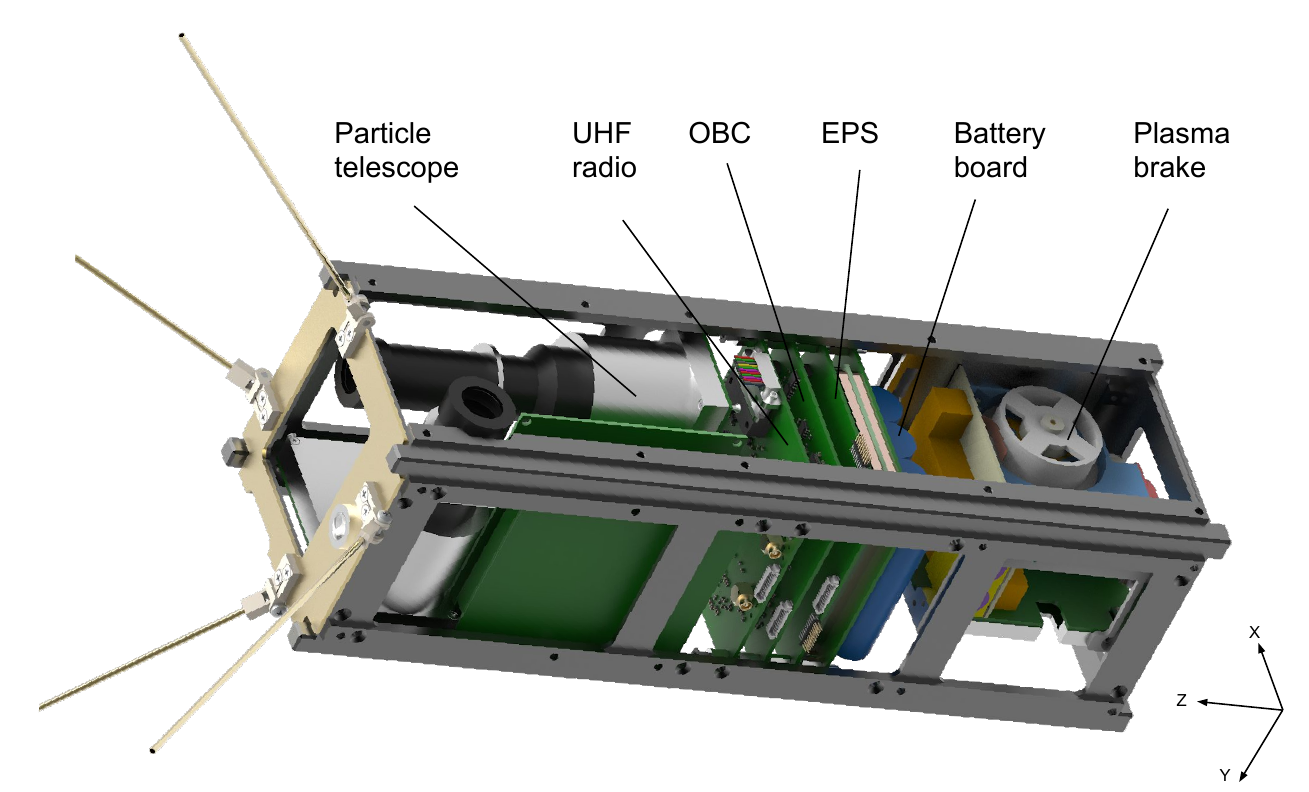}
\caption{Spacecraft structure including the subsystems without shielding.}
\label{fig:FS1_subsystems}
\end{figure}

The avionics stack consists of the Electrical Power System (EPS) for power collection, storage and distribution, Communication System for telemetry, On-Board Computer (OBC) for telemetry handling and mission and payload management and data storage, and Attitude Determination and Control System (ADCS) for maintaining the attitude modes during different operation phases. The mechanical structure satisfies dimensional limitations of the CubeSat standard and ensures modular configuration of the spacecraft's subsystems. The configuration of the platform is shown in Figure \ref{fig:FS1_subsystems}. The total mass budget is shown in Table \ref{tab:mass_budget}. 

\begin{table}
\centering
%\ravainio{Please note that PATE mass is now stabilized to 1.2 kg. We do not need a 20\% margin anymore, but cannot easily squeeze it back to 1.0 kg.}
\begin{tabular}{lccc}
Subsystem         & Planned mass (g)   & Mass with contingency (g)   & Fraction (\%)   \\
\hline
OBC               & 80                 & 96                          & 2             \\
EPS               & 926                & 1093                         & 23            \\
Magnetorquers     & 240                & 288                         & 6             \\
UHF Transceiver   & 80                 & 96                          & 2             \\
Antennas          & 50                 & 60                          & 1             \\
Structure         & 1033               & 1240                         & 26            \\
PATE              & 1000               & 1200                        & 26            \\
PB               & 600                & 660                         & 14            \\
\hline
Total             & 4009               & 4733                         & 100
\end{tabular}
\caption{FORESAIL-1 mass budget.}
\label{tab:mass_budget}
\end{table}

\subsection{Platform subsystems}
Each payload has different requirements for attitude. PATE needs to be oriented towards the Sun, while the detector with longer collimator scans the environment in the directions perpendicular to the Earth--Sun vector. The PB needs the spinning control for deploying and maintaining the tension of the tether. While the PB does not need specific pointing direction, its tether deployment introduces a mass distribution change that will require a proper control of the spacecraft angular momentum. 

The attitude determination and control system (ADCS) is divided into Attitude Determination System (ADS) and Attitude Control System (ACS). The ADS is equipped with gyroscope (L3GD20H), magnetometer (LIS3MDL), and in-house built sun sensors \citep{noman2017}. The angular velocity of the spacecraft is a crucial parameter for the payload attitude modes, thus gyroscope is necessary. Sun sensors are required for a precise sun pointing control, and magnetometers are needed for properly using with attitude control. The outputs will provide a full attitude information through using an unscented Kalman filter algorithm on all sensors.

The ACS changes the orientation of the spacecraft by using magnetorquers. There is a closed loop feedback, which ensures the maintenance of desired attitude by repeating the torque command until the desired orientation is achieved. The mission requirements for PATE and PB impose specific constraints on the attitude control. The spacecraft uses the following attitude control modes:
\begin{itemize}
\item Detumble mode: To stabilize the spacecraft after deployment
\end{itemize}
\begin{itemize}
\item Spin control mode: In order to deploy the PB tether, this mode spins up the spacecraft preferably with the spin axis being aligned with the earth pole. After deorbiting, spin down might be required during tether reel in if the tether does not break.
\end{itemize}
\begin{itemize}
\item Sun pointing mode: This mode continuously points Telescope 2 (at $-$Y axis) towards the sun while spinning in order for Telescope 1 (at Z-axis) to scan the sky.
\end{itemize}
 Magnetorquers are sufficient to provide all necessary control modes for the manipulation of the attitude and angular velocity. They are designed in-house, in form of air coils, so that they can be integrated to the solar panels (for the X and Y axes), and on a small factor magnetorquer driver board for the Z axis. All axes have two magnetorquers connected in parallel. The designed magnetorquers are driven through a custom-built coil driver to optimize either consumed power or the generated magnetic moment. The quantitative parameters of air core magnetorquers are given in Table \ref{tab:Platform_aircore}.

\begin{table}
\centering
\begin{tabular}{lcc}
Parameter          & X,Y axes  & Z axis   \\
\hline
Number of wire turns         & 184                & 952                       \\
Nominal current, mA   & 26.2                 & 19.3                       \\
Dipole moment, Am2               & 0.2                 & 0.2      \\
Power, mW & 51.7 & 56.5
                   \\
\hline\\
\end{tabular}
\caption{Quantitative parameters for the air core magneotorquers.}
\label{tab:Platform_aircore}
\end{table}

The EPS of the satellite consist of solar panels, power conditioning and power distribution \citep{ALI2014}, \citep{MUGHAL2014}.The solar panels are mounted on every long side of the satellite. The solar panels at other sides than where PATE is located consist of 7 solar cells connected in series, whereas the panel at the PATE side consists of 6 cells in order to provide space for the telescope. The power conditioning consists of switching buck converters to convert the incoming solar power to the battery voltage level. A perturb and observe based algorithm extracts the maximum power from the solar cells. Each subsystem houses a linear regulator to effectively convert the voltage down to a stable 3.3 V with a low ripple factor. The theoretical efficiency of the EPS is above 85 \%. All losses have been accounted for in the calculation of power budget. The maximum input power is 7 W, whereas the average power consumption in the nominal mode 3.7 W.

The on-board computer (OBC) is the satellite's main computer responsible for computational and data storage needs, running the ADCS algorithms, and storing the telemetry and housekeeping data logs. A safety-critical ARM Cortex-R4 based processor is selected as the central OBC of the spacecraft. The processor features 256 k data rapid-access memory and 3 MB of program flash. There are external flash memory devices also interfaced with the processor. To facilitate possible faults due to radiation, the OBC houses two cold redundant symmetric processors. Only one of the processors is active and powered. The arbiter switches the control to the redundant processor in case of failure. The OBC is responsible for operational work during the PB and PATE operations and collects all relevant telemetry data for downlink.

The UHF transceiver onboard the FORESAIL-1 consist of CC1125 transceiver with maximum output power of 15 dBm (30 mW); an external power amplifier RF5110G (gain: 31.5 dB, maximum output power 34.5 dBm) to amplify the power to desired 1.5-watt power in the transmit chain. In the receive chain, it consist of a Low Noise Amplifier and a bandpass filter.

\section{Ground Segment and Operations}

The primary ground station used for satellite operation is located in Aalto University Campus in Espoo, Finland. The ground station operates mainly as a radio amateur satellite station and has capabilities for operation on radio amateur UHF, VHF and S-bands. Due to its northern location the ground station has an average link time to polar orbit up to 90 minutes per day for all passes above the horizon. Ground station radio systems are built based on the Software Defined Radio (SDR) architecture which facilitates easy satellite-specific customization. The ground station infrastructure and equipment are designed and operated by students and serve also for educational purposes for the Aalto University. The Aalto University ground station also operates as the mission control centre. 

Typically, each satellite pass provides 10-20 min of link time between satellite and ground station.  The data rate requirements for FORESAIL-1 in science mode require the downlink data rate to be approximately 8 kbps. Since the data rate requirements are not stringent in order to accomplish both the missions, the ground station at Aalto University easily handles the data rate requirements.

\section{Data Products}

FORESAIL-1 data products are outlined in Table \ref{tab:data}. The Aalto University ground station is responsible for downlinking the L0/L1 data containing general spacecraft housekeeping data, PATE raw data, PB housekeeping and mission log information. These low level data products are shared forward using file-based web interface. For PATE data processing,  Level 1 raw data files combined to ADCS metadata and orbital information are used to produce calibrated and geolocated measurements of the particle fluxes. 

\begin{table}
\centering
\begin{tabular}{lll}
Provider & Data product & Details\\
\hline
Mission & List of data availability &  \\
Spacecraft & Orbital data, Attitude & Includes position and altitude required to \\
&& estimate the PB performance \\
PB & Electric current in the tether &\\
&	L0 telemetry stream &\\
&	L1 raw data &\\
&	L2 calibrated data &\\
&	L3 derived products & Plasma density\\
&& L2 and plasma density including position \\
PATE & Flux  &\\
&	L0 telemetry stream &\\
&	L1 raw data &\\
&	L2 calibrated data (fluxes) &\\
&	L3 derived products & L2 including location\\
&&	Precipitation maps (varying time resolutions)\\
&&	Angular distribution data (spin resolution)\\
&&	Event catalogues (mission duration)\\

\end{tabular}
\caption{FORESAIL-1 data products. PATE fluxes are given as function of energy, time and pointing azimuth.}
\label{tab:data}
\end{table}

These data products will be used to demonstrate that the PB de-orbits the satellite at a measurable amount, as a function of time. The PATE data will be used to infer the precipitation energy spectrum in time and place, addressing the science objectives. The science data are open for everyone, with the nominal rules-of-the-road typical in the field of space physics.

\section{Summary and Discussion}

The increasing number of small satellites launched into Earth's orbit raises timely concerns about the sustainable use of space. Small, rapidly built and launched satellites have a large future potential for scientific and commercial use. However, the satellites will become debris sooner rather than later, if they have poor radiation tolerance and if they are not de-orbited at the end of mission. The Finnish centre Of excellence in REsearch of SustAInabLe space (FORESAIL) funded by the Academy of Finland tackles this issue by focusing both on science of the near-Earth radiation environment and on novel technological solutions related to building more resilient instruments and debris removal. 

The first nanosatellite designed and built by the Centre of Excellence, FORESAIL-1, is a 3U cubesat operating at polar LEO orbit at and below 700 km which will produce energy-dependent pitch angle spectra of electrons and protons that precipitate from the RBs into the ionosphere. Further, it will measure energetic neutral atoms (ENAs) from the Sun and test the PB technology to lower the spacecraft altitude and manage its orbit in space.

Today, nanosatellites can address significant scientific questions. This requires focused measurements and innovative technological approaches, as well as coordination with the other spacecraft and facilities operating simultaneously. The FORESAIL-1 PATE instrument will make unprecedented and precise measurements of precipitating electrons and protons with high temporal resolution. It will be able to discriminate between electrons and protons, over a wide energy range (80 -- 800 keV for electrons and 0.3 -- 10 MeV for protons and neutral atoms). The large coverage of the polar cap at different orbital planes is achieved by operating the PB at the beginning of the mission, which sets the spacecraft orbital plane to drift in magnetic local time. This is the first time such a manoeuvre is attempted with a nanosatellite, and if successful, it will open new avenues for controlling the orbits of propellantless spacecraft, expanding their operational and observational ranges. 
 
FORESAIL-1 is targeted to make important advances in radiation belt physics. With simultaneous observations in the solar wind, magnetosphere and from the ground, FORESAIL-1 will allow quantifying the role of solar wind and outer magnetospheric driving, and the role of different plasma waves in the inner magnetosphere in the precipitation process. The primary science phase allows connecting precipitation signatures and mechanisms to geomagnetic activity levels and solar wind conditions. Together with the novel Vlasiator model, a global hybrid-Vlasov simulation \citep{Palmroth2018}, it will be possible to tie precipitation measurements to global processes in the outer magnetosphere for the first time, as Vlasiator begins its full six-dimensional operation before the launch of FORESAIL-1. The six dimensions refer to three dimensions in the ordinary space and three in the velocity space to describe the full particle distribution function, which is used to infer the energy spectrum and pitch angle in time. At the time of writing, Vlasiator allows already 2D electron precipitation calculations (see Fig. \ref{fig:Vlasiator_electrons}), which are in reasonable agreement with previous estimates \citep{Hardy1985}. Once the model is fully 6D, \textit{in situ} observations of electron precipitation can be directly compared to kinetic processes anywhere in the magnetosphere, without limitations as to whether a spacecraft traverses particular regions. This unprecedented plan will likely open new avenues in space physics.

Successful observations of ENAs from the Sun will allow estimating for the first time comprehensively estimating the suprathermal coronal ion population indirectly. This is the key knowledge for improved understanding of the charged particle acceleration processes at the CME-driven shock waves close to the Sun and of the CME energy budget. The solar ENA flux has been measured only once in a very fortuitous event \citep{Mewaldt2009}. Observing the ENA flux from the Sun on a regular basis as a function of time and solar activity is unprecedented.
 
The demonstration of the altitude manoeuvre of FORESAIL-1 will bring potential for future applications for the PB as a standard tool for removing satellites from their orbits. This is in particular a compelling solution considering the possibly upcoming regulations for including debris mitigation techniques in newly launched spacecraft.
 
FORESAIL-1 is at the forefront of scientific nanosatellites.  The advances we have made will be particularly important in demonstrating the usefulness of nanosatellites in making relevant physics and discovery measurements (ENA), whose spatio-temporal resolution could be brought to a new level using fleets of nanosatellites. Technological solutions of FORESAIL-1 have particularly far reaching impact for future debris removal solutions and orbit control. All these aspects are expected to pave the way for the sustainable use of space. 

\acknowledgments
The Finnish Centre of Excellence in Research of Sustainable Space, building and launching three FORESAIL missions is funded through the Academy of Finland with grant numbers 312351, 312390, 312358, 312357, and 312356. 

We acknowledge The European Research Council for Starting grant 200141-QuESpace, with which Vlasiator was developed, and Consolidator grant 682068-PRESTISSIMO awarded to further develop Vlasiator and use it for scientific investigations. We gratefully also acknowledge the Academy of Finland (grant numbers 138599, 267144, 309937, and 309939). We acknowledge the CSC -- IT Center for Science Grand Challenge grant for 2018, with which the Vlasiator simulation run was carried out. Vlasiator (http://www.physics.helsinki.fi/vlasiator/, \citep{wwwVlasiator}) is distributed under the GPL-2 open source license at https://github.com/fmihpc/vlasiator/ \citep{gitVlasiator}. Vlasiator uses a data structure developed in-house \citep{gitVlsv}, which is compatible with the VisIt visualization software \citep{HPV:VisIt} using a plugin available at the VLSV repository. The Analysator software (https://github.com/fmihpc/analysator/, \citep{gitAnalysator} was used to produce the presented figures. The run described here takes several terabytes of disk space and is kept in storage maintained within the CSC Ð IT Center for Science. Data presented in this paper can be accessed by following the data policy on the Vlasiator web site.

%\bibliography{bibligraphy.bib}

\end{document}